\documentclass[aps,prl,twocolumn,floatfix,nofootinbib,longbibliography]{revtex4}
\usepackage{amsmath}
\usepackage{graphicx}
\usepackage{amssymb}
\usepackage{amsmath}

\newcommand{\NOT}{\textsc{not}}
\newcommand{\CNOT}{controlled-\textsc{not}}

\begin{document}
\title{Efficient Hamiltonian programming in qubit arrays with nearest-neighbour couplings}
\author{Takahiro Tsunoda}
\author{Gaurav Bhole}
\author{Stephen A. Jones}
\author{Jonathan A. Jones}\email{jonathan.jones@physics.ox.ac.uk}
\author{Peter J. Leek}
\affiliation{Clarendon Laboratory, University of Oxford, Parks Road, Oxford OX1 3PU, United Kingdom}
\date{\today}

\begin{abstract}
We consider the problem of selectively controlling couplings in a practical quantum processor with always-on interactions that are diagonal in the computational basis, using sequences of local NOT gates. This methodology is well-known in NMR implementations, but previous approaches do not scale efficiently for the general fully-connected Hamiltonian, where the complexity of finding time-optimal solutions makes them only practical up to a few tens of qubits. Given the rapid growth in the number of qubits in cutting-edge quantum processors, it is of interest to investigate the applicability of this control scheme to much larger scale systems with realistic restrictions on connectivity. Here we present an efficient scheme to find near time-optimal solutions that can be applied to engineered qubit arrays with local connectivity for any number of qubits, indicating the potential for practical quantum computing in such systems.
\end{abstract}

\maketitle

\section{Introduction}
In gate-based quantum computation \cite{Barenco1995}, entangling interactions are generated by two-qubit logic gates, which are activated on demand. In order to turn the required two-qubit interactions on and off such gates are usually introduced by controllable external fields \cite{Kane1998}. However, such interactions can also be generated directly, without extra control hardware, when there are suitable coupling terms naturally present in the internal Hamiltonian \cite{Cory1998a,Zhou2002,Benjamin2003}.

To control such ``always-on'' couplings, decoupling and selective recoupling techniques were developed in the Nuclear Magnetic Resonance (NMR) community. Just as a spin echo refocuses the coherent dynamics of a spin-$1/2$ nucleus \cite{Hahn1950}, local \NOT\ gates on selected qubits in a multi-qubit system reverse the evolution of corresponding terms in the internal Hamiltonian  \cite{Jones1999,Leung2000}. By scheduling the timing and flip patterns properly, it is possible to rescale one-body and two-body terms to simulate any desired effective Hamiltonian \cite{Leung2002,Parra2020}.

Superconducting qubits commonly use gates generated on demand but can also be tuned to generate always-on two-qubit interactions \cite{Gambetta2017}. In fact always-on interactions are difficult to avoid in superconducting circuits due to their distributed microwave fields, and much effort is invested in mitigating these \cite{Chen2014,Mundada2019}. Such always-on couplings are present in capacitively-coupled transmon qubits induced by their relatively small anharmonicity \cite{DiCarlo2009}. Here, we leverage such a background Hamiltonian that contains single qubit interactions (offsets) $\Omega_j$   and two qubit interactions (couplings) $\omega_{jk}$ which are both diagonal in the computational basis,
\begin{equation}
    \mathcal{H}/\hbar = \sum_j \Omega_j \sigma_j^z/2 + \sum_{j<k} \omega_{jk} \sigma_j^z \sigma_k^z / 4.
\end{equation}
In principle superconducting quantum computers could be operated based on this NMR-like architecture, provided an efficient decoupling and recoupling scheme can be found to program a desired computation. We have previously described methods for finding time-optimal rescaling sequences in systems of up to around 20 qubits, and near-optimal sequences for around 100 qubits \cite{Bhole2020a}. However, these methods do not scale to larger systems, as the computational time required to find such solutions grows rapidly with the number of qubits.

The most general form of the Hamiltonian above will have all offsets and all possible couplings present, but couplings are generally only significant between nearby qubits. For example, superconducting qubits are often engineered in a square-lattice \cite{Arute2019} and can have couplings limited to only nearest or next-nearest neighbour qubits if circuits are well microwave engineered \cite{Spring2019}. Here we describe a method based on graph colouring for rapidly finding near-optimal sequences in these highly practical locally coupled systems. Remarkably we find that in this partially-coupled scenario, which is very realistic for large-scale superconducting circuits, the refocusing and rescaling of always-on couplings can be \textit{efficiently} programmed, requiring only linear time to design control sequences and linear number of control pulses.

We start by applying the refocusing scheme to the smallest possible case of a nearest-neighbour square lattice, and then generalise to  arbitrary lattice sizes. Initially we will assume the square lattice to be engineered with identical couplings between nearest neighbours and no long range couplings at all, but both of these restrictions will subsequently be partly relaxed.

\section{Square lattices}
\begin{figure}[tb]
\centering
\includegraphics{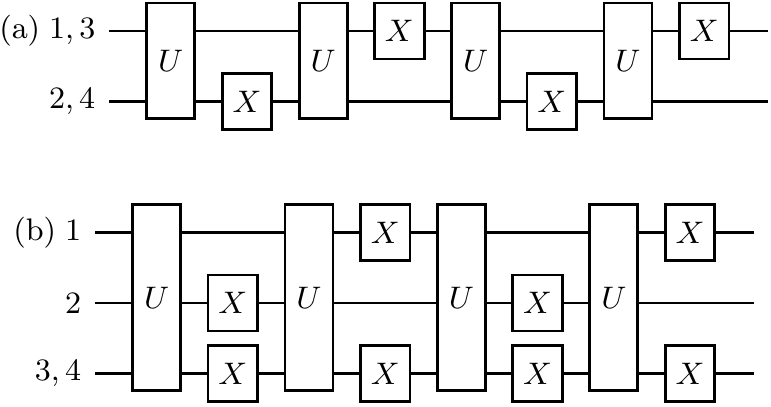}
\caption{Networks to (a) refocus all offsets and couplings in a square of four qubits with only nearest-neighbour couplings, and (b) keep only the coupling between qubits 3 and 4. Here $X$ indicates a \NOT\ gate while $U$ indicates evolution under the background Hamiltonian. Labels on the left indicate which qubits the \NOT\ gates are applied to.}
\label{fig:network}
\end{figure}

We begin with the smallest possible square lattice, containing four qubits with identical nearest neighbour couplings $\omega_{j,j+1}$ and $\omega_{14}$ only. Non nearest-neighbour pairs have no coupling, so $\omega_{13}=\omega_{24}=0$. We can refocus all the couplings in the system, as well as any offsets present, using the circuit of spin-echoes shown in Fig.~\ref{fig:network}(a). Here, the unitary $U$ describes the evolution of the system under the background Hamiltonian $\mathcal H$ given by the propagator $U = \exp{(-{\mathrm i}{\mathcal H}t/\hbar)}$, while the \NOT\ gates are represented by $X$. As there is no coupling to be refocused between qubits 1 and 3 it is possible to apply the same pattern of \NOT\ gates to them both, and similarly for qubits 2 and 4, so only two distinct patterns of \NOT\ gates are required.

This network can also be extended to retain a single coupling, as shown in Fig.~\ref{fig:network}(b). In this network the same pattern of \NOT\ gates is applied to qubits 3 and 4, and the corresponding coupling is retained, while the remaining couplings are refocused \cite{Jones1999}. These networks can be extended to retain \textit{any} pattern of couplings in a square lattice system.

\begin{figure}[tb]
\centering
\includegraphics{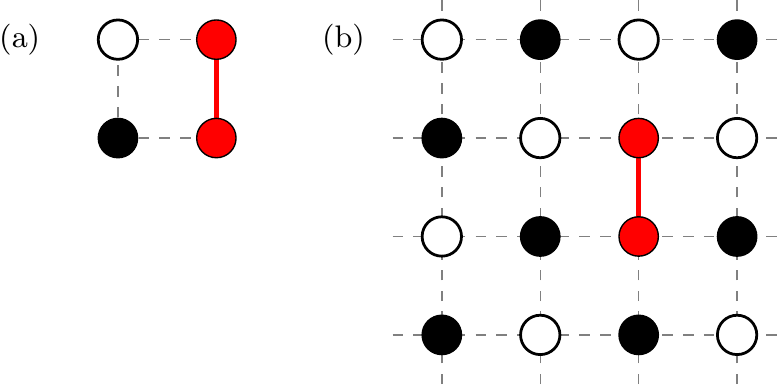}
\caption{(a) Retaining a nearest neighbour coupling in a square requires three colours: decoupled qubits are coloured black and white, while the coupled qubits are coloured red. Grey dashed lines show couplings which have been refocused. (b) The same result can be achieved in a larger array by colouring surrounding qubits alternately black and white.}
\label{fig:ABC}
\end{figure}

This extension works by colouring the square lattice, as described in \cite{Jones1999}. The system can be described as a noncomplete graph, with vertices corresponding to qubits and edges to couplings, with only some of the possible edges present.  The graph can be coloured by assigning a colour to each vertex, and is said to be properly coloured, corresponding to complete refocusing, if no two connected vertices are the same colour. Thus in the fully-decoupled square of four qubits we can colour qubits $1$ and $3$ black and qubits $2$ and $4$ white, while to retain a single coupling the pair of qubits involved must be assigned the same third colour, say red, as shown in Fig.~\ref{fig:ABC}(a). To implement a colouring, we use patterns of \NOT\ gates corresponding to distinct Walsh functions \cite{Jones1999,Bhole2020a} for each colour.

This colouring pattern can be tessellated across a lattice, as shown in Fig.~\ref{fig:ABC}(b), by colouring surrounding qubits  alternately black and white. Here we show a four-by-four patch containing sixteen qubits, which can be embedded in a larger lattice, retaining a single coupling while refocusing all the other interactions.
The required pulse sequence can be obtained by assigning black qubits $B$ to the first Walsh pattern, white qubits $W$ to the second, and the two red qubits $R$ to the third, to obtain the sequence
\begin{equation}
U\, X_\textsc{wr}\, U\, X_\textsc{BR}\, U\, X_\textsc{WR}\, U\, X_\textsc{BR},
\end{equation}
where $X_\textsc{WR}$ indicates that \NOT\ gates are applied to the white and red qubits, and so on.  Just like the sequence for the four-qubit system, this sequence requires only 4 time periods, but now requires $2q+4$ pulses for a system of $q$ qubits. The total time required to implement the network for a $\pi/2$ evolution, corresponding to a \CNOT\ gate, is $T=1/2J$, where the nearest-neighbour couplings are of size $2\pi J$. This is the same time as is needed for an isolated coupling, as the retained coupling evolves at full strength.

The approach above will refocus all single-qubit interactions, but it is simple to modify the $X$ gates in the network to implement single-qubit rotations directly.  This relies on the identity \cite{Morton2006a}
\begin{equation}
\pi_{\phi_2}\pi_{\phi_1}=2(\phi_2-\phi_1)_z,
\end{equation}
so that applying two $\pi$ rotations around axes in the $xy$-plane separated by an angle $\delta$ is equivalent to performing a $z$-rotation through an angle $2\delta$.  These rotations can be performed by modifying any pair of $X$ gates in any refocusing network, and as each qubit is controlled separately different rotations can be applied to different qubits at no cost in time or pulse count.

\section{Parallel gates}
\begin{figure}[tb]
\centering
\includegraphics{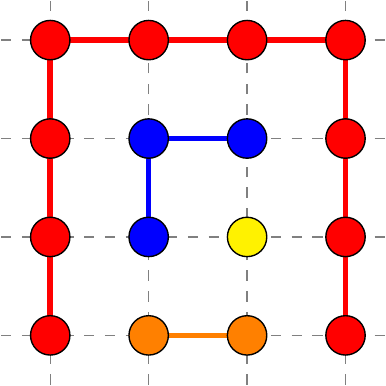}
\caption{Retaining a more complex pattern of couplings in a square lattice: this pattern requires four colours.}
\label{fig:fourcolour}
\end{figure}

If it is desired to retain several different coupling interactions then this can be achieved most simply by applying such patterns back to back, changing the colouring at each stage, but it is more efficient to as far as possible perform evolutions in parallel.  For simple cases this can be achieved as shown in Fig.~\ref{fig:fourcolour}.
Qubits which are part of the same coupling island, that is qubits which are connected either directly or indirectly by retained couplings, have been coloured the same colour. (Note that the single yellow qubit forms an island on its own.)

This simple approach will only be successful when, as here, all the couplings which could appear in an island are retained.  In such cases this simple colouring strategy will retain all the desired couplings while refocusing all the couplings between islands, reproducing the desired pattern in one go. As before, the total time required is just $T=1/2J$, the same as for an isolated coupling.

It might appear that this approach would require the number of colours to equal the number of islands, but in fact it is never necessary to use more than four, since islands which are completely disconnected (that is, islands which are not connected by couplings in the underlying Hamiltonian) can be safely coloured the same. By the four-colour-map theorem this will never require more than four colours \cite{Appel1977}. These four colours can be assigned to four Walsh functions \cite{Bhole2020a}, requiring $T$ to be divided up into eight equal time periods. Assigning the most common colours to $W_1$ and $W_2$, which both require two pulses, and the two rarer colours to $W_3$ and $W_4$, which require four pulses, means that the total number of pulses required in a system of $q$ qubits lies between $2q$ and $3q$.

\begin{figure}[tb]
\centering
\includegraphics{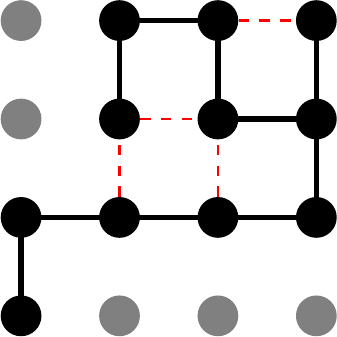}
\caption{A target pattern of couplings which cannot be implemented with a single colouring pattern. Although the grey qubits are easily decoupled from each other and from the black island, it is impossible to retain all the desired black couplings without also retaining the undesired couplings shown as dashed red lines.}
\label{fig:target}
\end{figure}

Although this method will work for some simple target patterns, it will not work in general, since many target patterns have one or more missing couplings. Consider, for example, the pattern shown in Fig.~\ref{fig:target}, where the black couplings must be retained, and the dashed red couplings must not be retained although they connect qubits within the main island.
Any colouring which implements all the black couplings that must be retained will also implement the unwanted red couplings, and so this target pattern cannot be implemented with any single colouring pattern.

\section{Multiple colourings}
\begin{figure}[tb]
\centering
\includegraphics{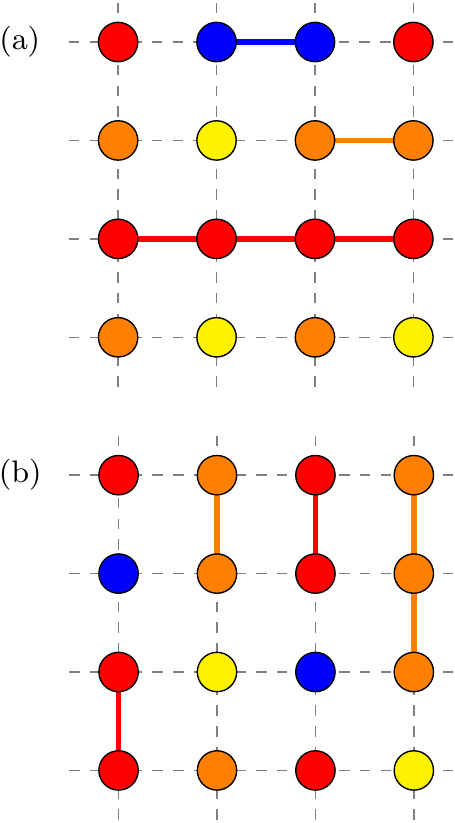}
\caption{A solution implementing the target couplings in Fig.~\ref{fig:target} using two sequential colouring patterns, (a) and (b).}
\label{fig:pair}
\end{figure}

There is, however, a simple method for achieving \textit{any} target pattern with two sequential colourings, each using four colours. As an example a pair of colourings which implements the target couplings in Fig.~\ref{fig:target} is shown in Fig.~\ref{fig:pair}.
Pattern (a) assigns two colours to the odd-numbered rows and two more to the even numbered rows, thus ensuring horizontal couplings can be controlled while all vertical couplings will be refocused. Along a row, the colour of the first qubit is arbitrary, but the following qubit must be the same colour if the corresponding coupling is to be retained, and the other colour if it is to be refocused. Pattern (b) assigns colours to columns instead of rows to control vertical couplings in an analogous way.

As any subset of horizontal couplings can be selected by the first colouring, and any subset of vertical couplings can be selected by the second, any target pattern at all can be implemented in this way, in a total time $2T$. This total time will be divided into 16 equal time periods, separated by no more than $6q$ individual pulses.

It is useful to compare this result with timings found by our previous algorithm \cite{Bhole2020a} based on linear programming with an exhaustive basis set, which is guaranteed to find a true minimum-time solution. The computational complexity of this algorithm renders it impractical for large arrays, but for four-by-four arrays with nearest-neighbour couplings it is perfectly practical. We have analysed a very large number of randomly chosen targets in four-by-four arrays, and in every case the optimal solution required a total time of either $T$ (in cases with no missing couplings, so a single colouring is possible) or $1.5\,T$ (in cases where this is not possible).

The implementation time required for our colouring based networks, $2T$, is slightly longer than the absolute minimum required, but these colourings are far easier to design, with a computation time scaling only \textit{linearly} with the total number of qubits, and so can be applied to systems any number of qubits. The number of \NOT\ gates required is also greatly reduced, from $O(q^2)$ for linear programming based solutions to $O(q)$. 

\section{Next-nearest neighbours}
Until now we have assumed that only nearest-neighbour interactions are important, with all others being too small to matter.  In practice, real physical systems are likely to also have non-zero couplings at longer range. It is reasonable to expect such couplings to drop with distance in well-engineered devices, making the next-nearest neighbour interactions, across the diagonals of the square array, the most significant. In the original two-colour refocusing scheme qubits connected by next-nearest neighbour interactions will be the same colour, both black or both white, and so these interactions will be retained rather than refocused, leading to a significant error in the final gate implementation.

This is not a problem with the general four-colour scheme, as the use of different colours for alternate rows and columns guarantees that diagonally connected qubits will be of different colours. Thus these networks automatically suppress any unwanted diagonal couplings, which can therefore simply be ignored. Suppressing even longer range couplings is more complicated, but can be achieved using larger numbers of colouring patterns with more colours used in each pattern.

\section{Different evolution times}
The parallel gates approach, however, hides a further important assumption: it is not sufficient to retain two different couplings if they are required to evolve for different times, either because different couplings must evolve through different angles, or because apparently equivalent couplings will have different strengths, and so will need different evolution times to achieve the same angle.

It might seem necessary to apply such gates using different echo sequences, but in fact they can be partly combined.  Consider two couplings in the same group, where one requires evolution for a total time $T_A$ and the second for a time $T_B$, with $T_B>T_A$.  The naive approach is to use two different colouring patterns, one implementing the first coupling for time $T_A$, and another implementing the second coupling for $T_B$. In fact these periods can be carried out partly in parallel: during the first period, which lasts for time $T_A$, \textit{both} couplings are retained, while for the second period, which lasts for time $T_B-T_A$, only the second coupling is retained.  Thus both couplings can be carried out in a total time $T_B$, and the method generalises for any number of distinct couplings strengths.

It follows that \textit{any} pattern of couplings can be achieved in an evolution time equal to the sum of the longest evolution times required for horizontal and vertical couplings, which itself is no more than $2T_{\rm max}$, where $T_{\rm max}=1/2J_{\rm min}$ is the evolution time required for the slowest  gate.
However the resulting sequences will contain $O(q)$ evolution times and $O(q^2)$ pulses, while designing such sequences requires sorting the evolution times into ascending order, with computational time complexity $O(q\log q)$. They will also be impractical to implement experimentally, as the differences between very similar times may be smaller than the clock resolution.

Rather than implementing a very large number of distinct evolution times precisely, it makes more sense to use a much smaller number of evolutions to approximate all the desired times. This is most easily achieved by dividing the range of evolution times by successive powers of two, in effect encoding each evolution time as a binary number. Using $k$ different evolution times results in $k$-bit precision, with an exponential increase in precision with a linear increase in the number of evolution times used. By accepting a degree of approximation one can reduce the time complexity from $O(q \log q)$ to $O(k q)$, and reduce the pulse count from $O(q^2)$ to $O(k q)$, where the constant $k$ depends on the accuracy required. For example, using 20 distinct delays will allow angles to be approximated to a precision better than $10^{-6}$.

\section{Conclusions}
The colouring techniques described here allow the efficient control of interactions in qubit arrays of arbitrary size, provided the couplings are constrained to be local. While the resulting pulse sequences take slightly longer to implement than the absolute minimum time required, the computational time is vastly reduced, from $O(4^q)$ for exhaustive linear programming, or $O(q^6)$ for RROS \cite{Bhole2020a}, right down to $O(q)$, rendering them practical in systems with thousands or even millions of qubits. The number of control pulses is also greatly reduced, from $O(q^2)$ to $O(q)$, thus reducing implementation errors. The method can handle unwanted next-nearest neighbour couplings, and is easily extended in a scalable way to systems with variable coupling strengths or evolution angles.

Although it is not yet common to use always-on couplings as a computational resource in superconducting qubits, such implementations might fit well into the context of Near-term Intermediate Scale Quantum (NISQ) applications \cite{Preskill2018}. For example, the native interaction can be used for hardware-efficient implementations of the variational quantum eigensolver (VQE) \cite{Kandala2017} and quantum approximate optimisation algorithms (QAOA) \cite{Farhi2014}. Since entangling gates are realized without external fields, there is less need to calibrate the control hardware, as the calibration problem can be focused only on local control. Furthermore, quantum error mitigation methods might allow slow native interactions to output useful results by extended time evolution \cite{Temme2017, LiY2017, Kandala2019}. Efficient refocusing and rescaling methods, such as those described here, will be essential to extend experiments to the next generation of devices.

\begin{acknowledgments}
TT is supported by the Masason Foundation and the Nakajima Foundation. GB is supported by a Felix Scholarship. PL acknowledges support from the EPSRC under grant EP/T001062/1.
\end{acknowledgments}
\bibliography{hrefs}
\end{document}